\documentclass[preprint,onecolumn,10pt]{revtex4-1}
\usepackage{graphicx}
\usepackage{color}
\usepackage{amsmath}

\begin{document}

\title{Detailed analysis of the performance of the modified Becke-Johnson potential}
\author{J. A. Camargo-Mart\'inez}
\affiliation{Departamento de F\'isica, CINVESTAV-IPN, Av. IPN 2508, 07360 M\'exico}
\author{R. Baquero}
\affiliation{Departamento de F\'isica, CINVESTAV-IPN, Av. IPN 2508, 07360 M\'exico}

\begin{abstract}

Very recently, in the 2011 version of the Wien2K code, the long standing shortcome of the codes based on Density Functional Theory, namely, its impossibility to
account for the experimental band gap value of semiconductors, was overcome. The novelty is the introduction of a new exchange and correlation potential, the 
modified Becke-Johnson potential (mBJLDA). In this paper, we report our detailed analysis of this recent work. We calculated using this code, the band structure 
of forty one semiconductors and found an important improvement in the overall agreement with experiment as Tran and Blaha [{\em Phys. Rev. Lett.} 102, 226401 (2009)]
did before for a more reduced set of semiconductors. We find, nevertheless, within this enhanced set, that the deviation from the experimental gap value can reach 
even much more than 20\%, in some cases. Furthermore, since there is no exchange and correlation energy term from which the mBJLDA potential can be deduced, a 
direct optimization procedure to get the lattice parameter in a consistent way is not possible as in the usual theory. These authors suggest that a LDA or a GGA 
optimization procedure is used previous to a band structure calculation and the resulting lattice parameter introduced into the 2011 code. This choice is important 
since small percentage differences in the lattice parameter can give rise to quite higher percentage deviations from experiment in the predicted band gap value. 
We found that by using the average of the two lattice parameters (LDA and GGA) a better agreement with the band gap experimental value is systematically obtained. 
As a rule, the LDA optimization underestimates the lattice parameter while the GGA one overestimates it. Also we found that using the experimental lattice 
parameter instead, surprisingly high deviations of the predicted band gap value from experiment, occur. This is an odd result since, in general, the quality of the 
LDA and GGA obtained lattice parameters are judged to be as good as their proximity to the experimental lattice parameter value. This judgment implies the idea 
that the best result for the predicted band gap value is obtained when the closest-to-experiment lattice parameter is used. On the other hand, the band structure 
calculated with the mBJLDA potential seems, at first sight, a simple rigid displacement of the conduction bands towards higher energies. A closer look reveals that, 
in some cases, important differences occur that might not be negligible in certain systems containing a semiconductor as it might happen at interfaces. So, in some 
systems containing a semiconductor, neither the direct use of the Wien2k previous version nor its use with a rigid displacement of the conduction bands added so as 
to reproduce the band gap value, are totally reliable. The overall implementation of the calculation of the band structure of semiconductors with the Wien2k code 
using this new potential is quite empirical although it mimics well the results obtained by other methods as the GW approximation which give better results and are 
theoretically well founded. We conclude that, in spite of the very important improvement in the band gap agreement with experiment using 
the mBJLDA potential, there are issues that point to the fact that this problem is not yet totally closed.

\end{abstract}
\date{\today}

\pacs{71.15.Mb;71.20.Mq;71.20.Nr}
\keywords{Wien2K; mBJLDA potential; EOS Murnaghan; semiconductors; insulator; band structure; band gap.}


\maketitle

\section{Introduction \label{introduction}}

Wien2k~\cite{1} has been one of the most used codes to calculate the band structure of solids since long ago. In 2011, the long standing shortcome, i.e., its 
impossibility to reproduce the experimental gap of semiconductors, has been approximately overcome in the Wien2k 2011 version~\cite{2}. This code is based on the Full 
Potential (Linearized) augmented plane waves and local orbitals [FP-(L)APW+lo]~\cite{3} where a new exchange and correlation potential was introduced. 
The result is a very remarkable improvement with respect to the previous version. Some cases worth noting remain as an exception but 
the overall agreement is good. While computing the gap of a certain number of semiconductors (41), we found that optimization plays a role since small 
differences in the lattice parameter can give, for the same semiconductor, noticeable differences in the gap.
It was observed that small changes in the lattice parameter between 0.02\% and 1.0\%, can induce differences between 0.5\% and 6.5\% 
(see text below). The optimization procedure gives the value of the lattice parameter that is consistent with the minimum of the energy vs. volume curve. As such, 
it depends on the interchange and correlation energy, $E_{xc}$, used in the calculation. In principle, it appears that the consistent procedure should be to use in 
the optimization procedure from which the lattice parameter is obtained, the exchange and correlation energy functional from which the exchange and 
correlation potential to be used in the electronic band structure is deduced. The previous version of the Wien2K code uses either LDA or GGA in this way but as it is well 
known, neither formulation reproduces accurately enough the gap of semiconductors. The exchange and correlation potential, the modified Becke-Johnson potential 
(mBJLDA), does a much better job. Blaha et al.~\cite{2} got the best results when the lattice parameter is first obtained from a LDA or a GGA optimization followed 
by a band structure calculation using the new mBJLDA exchange and correlation potential. This way of performing the calculation is necessary since it is impossible 
to obtain an expression for a functional $E_{xc}$ such that the mBJLDA potential is $V_{mBJ}=\delta E_{xc}[\rho]/\delta \rho$, as in the usual theory. In this work, 
we analyse the results obtained with the mBJLDA to pinpoint to its assets and to its odds in more detail. We conclude that even though the overall
agreement with experiment of the band gap value obtained with the Wien2k 2011 constitutes a noticeable improvement, there are 
some cases and details that point to the fact that the problem might not be really totally closed.

\section{Density Functional Theory}
Several codes were developed based on Density Functional Theory (DFT) which became the most used, precise and practical way to calculate the band structure of 
solids. The development of practical approximations to the exchange and correlation energy functional lead to a remarkable degree of accuracy to describe even 
complicated metallic systems. At the basis of DFT is the Hohenberg-Khon theorem~\cite{4} which shows that the knowledge of the density of the ground state is 
equivalent to the one of the wave function itself. The density of states of the real ground state many body system is equal to the one calculated from the 
solution of the Khon-Sham equations~\cite{5},
\begin{equation}
\label{eq:eq1}
[T+V_{H}+V_{xc}+V_{ext}]\varphi_{i} (r)= \varepsilon_{i}\varphi_{i}
\end{equation}
where the density is calculated taking into account the occupied states only. In eq.(\ref{eq:eq1}), $T$ is the kinetic energy operator, $V_{H}$ is the Hartree 
potential, $V_{ext}$ is the external potential and $V_{xc}$ is the exchange and correlation potential which is calculated from the exchange and correlation 
energy functional; $V_{xc}=\delta E_{xc}[\rho]/\delta \rho$. To solve the Khon-Sham equations (\ref{eq:eq1}), an explicit expression for $E_{xc}[\rho]$ is needed. 
The exact expression is unknown since it includes all kind of correlations between all the particles in the system. So an approximation is needed. The first and 
best known approximation is the Local Density Approximation, LDA~\cite{6}, which was followed by the Generalized Gradient Approximation (GGA)~\cite{7} and the 
meta-GGA~\cite{7} among others. These potentials reproduce rather well the band structure of even complicated metallic systems but fail to reproduce the gap of 
semiconductors. As a possible empirical solution to this problem, Blaha et al.~\cite{2} have reported the mBJLDA potential which is a modification of the exchange and correlation potential 
of Becke and Johnson (BJ)~\cite{8}. The new potential reproduces the experimental gap of semiconductors with accuracy several orders of magnitude better than 
the previous version of the Wien2K code using either the LDA or the GGA. The mBJLDA potential is 
\begin{equation}
\label{eq:eq2}
V_{x,\sigma}^{MBJ}(r)=cV_{x,\sigma}^{BR}(r)+(3c-2)\frac{1}{\pi}\sqrt{ \frac{5}{12}}\sqrt{ \frac{ 2t_{ \sigma}(r)}{\rho_\sigma(r)}}
\end{equation}
Where $\rho_\sigma(r)$ is the spin dependent density of states, $t_\sigma(r)$ is the kinetic energy density and $V_{x,\sigma}^{BJ}(r)$ is the Becke-Roussel
potential (\emph{BR})~\cite{9}. The c stands for
\begin{equation}
\label{eq:eq3}
c=\alpha + \beta \left(\frac{1}{V_{cell}}\int{d^3 r\frac{\mid \nabla \rho(r)\mid}{\rho(r)}}\right)^{1/2}
\end{equation}
$\alpha$ and $\beta$ are free parameters. The \emph{Wien2k 2011} code defines $\alpha=-0.012$ and $\beta=1.023$ $Bohr^{1/2}$. These values are general but 
certainly fixed experimenting with several cases. No expression for the exchange and correlation energy is given and therefore no total energy functional 
is really possible. It is in this sense that the formulation might be seen as an empirical model in spite of its universality. We next explore 
further some more details of it.

\section{The optimization procedure}
The optimization procedure in the usual theory, uses the proposed energy functional to get the structural ground-state parameters of the solid at T=0 K. These
parameters are then used in a further step to get the band structure in a consisted way. It is in this sence that the exact way used for the optimization 
procedure is important.
The Wien2k code allows the calculation of the equilibrium structural properties of the system, the minimum of the total energy, ($E_{0}$), the Bulk modulus, 
($B_0$), its derivative with respect to pressure, ($B_0'$), and the equilibrium volume at zero pressure, $V_0$, by fitting the data to an equation of state (EOS). 
The code uses either the EOS by Murnaghan~\cite{10} or the one by Birch-Murnaghan~\cite{11} or else the so-called EOS2~\cite{12}. We used the first one, 
\begin{equation}
\label{eq:eq4}
 E(V)=E_0+\frac{B_0V}{B_0'}\left[\frac{1}{(B_0' -1)}\left(\frac{V_0}{V}\right)^{B_0'}+1\right]-\frac{B_0V_0}{(B_0'-1)}
\end{equation}
\begin{equation}
\label{eq:eq5}
 P(V)=\frac{B_0}{B_0'}\left[\left(\frac{V_0}{V}\right)^{B_0'}-1\right]
\end{equation}
and performed the calculation for some semiconductors (Si, Ge, AlAs, SiC, BP, AlP, GaN, GaAs, LiF, MgO and BN), using the LDA and GGA approximations. 

\begin{table}[ht]
\caption{\label{tab:Tabla1}The Bulk modulus, ($B_0$), in GPa and its derivative, ($B_0'$), as obtained from the Murghanan fit. 
The crystal structure is shown in parenthesis. Experimental values ​​were taken from references~\cite{13,14,15,16,17,18,19,20,21,22,23}.}
\begin{tabular*}{0.6\textwidth}{@{\extracolsep{\fill}} l  c c c c c c }\hline\hline
          & \multicolumn{2}{c}{LDA}  & \multicolumn{2}{c}{GGA} &  \multicolumn{2}{c}{Expt.}\\
\hline
Solid     & $B_0 $   & $B_0'$& $B_0$    & $B_0'$&  $B_0$    & $B_0'$ \\
\hline
C(A1)     & 434.34   & 3.69  & 432.52   & 3.99  &     442   &  4.03 \\
Si(A1)    & 92.96    & 4.35  &  87.71   & 4.23  &    97.82  &  4.09  \\
Ge(A1)    & 73.18    & 4.98  &  62.45   & 3.93  &    75.80  &  4.55  \\
MgO(B1)   & 175.52   & 4.61  &  149.89  & 4.18  &    160.00 &  4.15  \\
LiF(B3)   & 69.29    & 4.91  &  68.07   & 4.22  &    62.00  &  5.14  \\
AlAs(B3)  & 73.34    & 4.78  &  67.29   & 4.53  &    78.10  &  -     \\
SiC(B3)   & 233.85   & 4.06  &  214.95  & 4.11  &    224.00 &  4.1  \\
BP(B3)    & 175.12   & 3.85  &  160.91  & 3.84  &    173.00 &  3.76 \\
AlP(B3)   & 89.70    & 4.17  &  82.15   & 3.97  &    86.00  &  3.99  \\
BN(B3)    & 401.27   & 3.71  &  370.10  & 3.66  &    369.00 &  4.00  \\
GaN(B3)   & 202.45   & 3.28  & 177.39   & 4.11  &    188.00 &  3.2   \\
GaAs(B3)  & 75.15    & 3.97  & 67.14    & 4.03  &    74.66  &  4.6   \\
\hline\hline
\end{tabular*}
\end{table}

In the Table~\ref{tab:Tabla1}, we present these results first for the Bulk modulus ($B_0$) and its derivative with respect to pressure ($B_0'$). We can see
that when calculated with the LDA, the deviations from experiment of the Bulk modulus and its derivative oscillate roughly between 1-12\%; with the GGA between 
4-18\%. The values for the Bulk modulus as compared to experiment are overestimated in all cases but three (Si, Ge, AlAs) when an LDA optimization is used. For a 
GGA one, all are underestimated but two (LiF, BN). The GGA optimization gives a better agreement with experiment in five of the eleven semiconductors considered 
(MgO, LiF, AlP, BN, GaN). Looking at the derivative of the Bulk modulus with respect to pressure (seven experimental results reported), the GGA optimization 
gives a better agreement with experiment in four cases (Si, Ge, MgO, GaN).

One the problem with the mBJLDA potential is that since no exchange and correlation energy functional, $E_{xc}$, is defined within this formulation, no consistent
optimization procedure is possible. This shortcome is the result of the empirical character of the mBJLDA potential. As a somehow empirical and inconsistent 
solution to this shortcome, Tran and Blaha~\cite{2} suggest to start with an GGA (or LDA) optimization and to introduce the lattice parameter value obtained 
into the band structure calculation code that uses the mBJLDA potential.

\begin{table}[h]
\caption{\label{tab:Tabla2}Lattice parameters, ($a$), obtained from the Murnaghan fit. The $a_{Avg}$ values ​​are the average 
value between $a_{LDA}$ and $a_{ GGA}$. The experimental data, at low temperature (LT) and room temperature (RT), are from refs.~\cite{23,24,25}.}
\begin{tabular*}{0.5\textwidth}{@{\extracolsep{\fill}} l  c c c c c c}\hline\hline
        &  \multicolumn{6}{c}{Lattice parameters $a$[\AA]} \\\hline
Solid     & &$a_{LDA}$ & $a_{GGA}$    & $a_{Avg}$ & $a^{ Expt.}_{LT}$ &$a^{Expt.}_{RT}$ \\ 
\hline
C(A1)     & &  3.5339  &  3.5731      &  3.5535   & 3.5667* & 3.5668 \\
Si(A1)    & &  5.4073  &  5.4738      &  5.4406   & 5.4298* & 5.4310 \\
Ge(A1)    & &  5.6269  &  5.7586      &  5.6928   & 5.6524* & 5.6579 \\
MgO(B1)   & &  4.1635  &  4.2569      &  4.2102   & 4.2052*$^\text{a}$ & 4.2110 \\
LiF(B3)   & &  3.9152  &  4.0710      &  3.9931   &    -    & 4.0300 \\
AlAs(B3)  & &  5.6329  &  5.7304      &  5.6817   & 5.6605* & 5.6614 \\
SiC(B3)   & &  4.3378  &  4.3944      &  4.3661   & 4.3585$_{\text {\tiny 3 K}}^\text{b}$ & 4.3596 \\
BP(B3)    & &  4.4937  &  4.5510      &  4.5224   &    -    & 4.5383 \\
AlP(B3)   & &  5.4371  &  5.5110      &  5.4741   &    -    & 5.4635 \\
BN(B3)    & &  3.5831  &  3.6268      &  3.6050   &    -    & 3.6155 \\
GaN(B3)   & &  4.4663  &  4.5626      &  4.5145   &    -    & 4.5230 \\
GaAs(B3)  & &  5.6050  &  5.7401      &  5.6726   & 5.6523* & 5.6533 \\
BAs(B3)   & &  4.7314  &  4.8085      &  4.7700   &    -    & 4.777 \\
InP(B3)   & &  5.8239  &  5.9535      &  5.8887   & 5.8657$_{\text {\tiny 4.8 K}}$& 5.87875 \\
AlSb(B3)  & &  6.1084  &  6.2209      &  6.1647   &     -   & 6.1355 \\
GaSb(B3)  & &  6.0449  &  6.2086      &  6.1268   &     -   & 6.09593 \\
GaP(B3)   & & 5.3912   &  5.5084      &  5.4498   & 5.4469* & 5.4508 \\
InAs(B3)  & &  6.0185  &  6.1804      &  6.0995   &    -    & 6.0584 \\
InSb(B3)  & &  6.4497  &  6.6363      &  6.5430   &    -    & 6.4794 \\
CdS(B3)   & & 5.7636   &  5.9304      &  5.8470   &    -    &  5.8320 \\
CdTe(B3)  & &  6.4166  &  6.6198      &  6.5182   & 6.4675$^\text{c}$  &  6.4810 \\
CdSe(B3)  & &  6.0169  &  6.1995      &  6.1082   &    -    & 6.0770 \\
ZnS(B3)   & & 5.3083   &  5.4524      &  5.3804   &    -    & 5.4102 \\
ZnSe(B3)  & & 5.5826   &  5.7401      &  5.6614   &    -    & 5.6680 \\
ZnTe(B3)  & & 6.0129   &  6.1854      &  6.0992   &    -    & 6.0890 \\ 
MgS(B1)   & & 5.1358   &  5.2342      &  5.1850   &    -    & 5.2030 \\
MgS(B3)   & & 5.5985   &  5.7018      &  5.6502   &    -    & 5.6220 \\
MgSe(B1)  & & 5.3993   &  5.5129      &  5.4561   &    -    & 5.4600 \\
MgTe(B3)  & & 6.3836   &  6.5187      &  6.4512   &    -    & 6.4230 \\
BaS(B1)   & &  6.2743  &  6.4313      &  6.3528   &    -    & 6.3890$^\text{d}$ \\
BaSe(B1)  & &  6.4792  &  6.6604      &  6.5698   &    -    & 6.5950$^\text{d}$ \\
BaTe(B1)  & & 6.8647   &  7.0680      &  6.9664   &    -    & 7.0070$^\text{d}$ \\
CaO(B1)   & & 4.7177   &  4.8346      &  4.7762   &    -    & 4.8110 \\
CuCl(B3)  & &5.2101    &  5.4403      &  5.3252   & 5.4093*$^\text{a}$ & 5.4202 \\
CuBr(B3)  & & 5.5264   &  5.7391      &  5.6328   & 5.6764* & 5.6900 \\
AgF(B1)   & &  4.7946  &  5.0224      &  4.9085   &    -    & 4.9360 \\
AgI(B1)   & &  6.3656  &  6.6416      &  6.5036   &    -    & 6.4990 \\
GaN(B4)   &a&  3.2027  &  3.2051      &  3.2039   &    -    & 3.1980 \\
          &c& 5.1408   &  5.1330      &  5.1369   &    -    & 5.1850 \\ 
InN(B4)   &a& 3.5715   &  3.5751      &  3.5733   &    -    & 3.5480 \\
          &c& 5.6845   &  5.6729      &  5.6787   &    -    & 5.7600 \\
AlN(B4)   &a& 3.1200   &  3.1211      &  3.1206   & 3.1113*$^\text{e}$ & 3.1120 \\
          &c& 4.9565   &  4.9529      &  4.9547   & 4.9793*$^\text{e}$ & 4.9820 \\
ZnO(B4)   &a& 3.2675   &  3.2604      &  3.2640   & 3.2482$_{\text {\tiny 4.2 K}}^\text{f}$ & 3.2500 \\
          &c& 5.1736   &  5.1653      &  5.1695   & 5.2040$_{\text {\tiny 4.2 K}}^\text{f}$& 5.2070 \\
\hline\hline
\multicolumn{6}{l}{*Extrapolated values. $^\text{a}$Ref.~\cite{26}, $^\text{b}$Ref.~\cite{27},}\\
\multicolumn{6}{l}{  $^\text{c}$Ref.~\cite{28}, $^\text{d}$Ref.~\cite{29}, $^\text{e}$Ref.~\cite{30}, $^\text{f}$Ref.~\cite{31}.}\\
\end{tabular*}
\end{table}
\newpage
\clearpage

In Table \ref{tab:Tabla2}, we present the lattice parameter, $a$, calculated with LDA, GGA, its average value, $a_{Avg}$, ($a_{Avg}=(a_{LDA}+a_{GGA})/2$)
and the experimental reports at low temperature (LT) and at room temperature (RT). Notice that the LDA gives rise to deviations of the lattice parameter 
that are always below the experimental value (except for the values ​​of $a$ in hexagonal structures) while exactly the contrary arises with the GGA 
(except for the values ​​of $c$ in hexagonal structures). GGA results in a shift towards higher values of the equilibrium volume which might be 
thought as a kind of relaxation due to the use of derivatives of the local density in the functional. The experimental lattice parameters were taken from 
refs.~\cite{23,24,25}. In Table \ref{tab:Tabla2}, we first compare the lattice parameter obtained with an LDA optimization to experimental data taken at 
low temperature (LT) and at room temperature (RT). In the first case, we find, in general, small differences (0.6-1\%), except for CuBr and CuCl (2.6 and 3.7, 
respectively). In the second case, the differences are usually below 1.3\% but can rise up to 3.9\% as in CuCl. If we use the GGA approximation the differences 
are slightly higher in general. 

We will show below that the use of the average lattice parameter, $a_{Avg}$, results in a gap value that is always in better agreement with experiment. We give 
an error statistics analysis of this result in Table \ref{tab:Tabla3} to emphasise its validity. To obtain $a_{Avg}$ has a low computational extra cost.

\begin{table}[h]
\caption{\label{tab:Tabla3} Lattice parameter error statistics for compounds of Table \ref{tab:Tabla2}, in \AA.}
\begin{tabular*}{0.4\textwidth}{@{\extracolsep{\fill}} c c c c}\hline\hline
\multicolumn{4}{l}{{\small Error relative to LT experiments.}}\\\hline
               & $a_{LDA}$    & $a_{GGA}$   & $a_{Avg}$ \\\hline
ME$^\text{a}$  & -0.046 & 0.047 & 0.00041   \\
MAE$^\text{b}$ & 0.050  & 0.055 & 0.025     \\
SD$^\text{c}$  & 0.046  & 0.049 & 0.034     \\
\hline
\multicolumn{4}{l}{{\small Error relative to RT experiments.}}\\\hline
               & $a_{LDA}$    & $a_{GGA}$   & $a_{Avg}$ \\\hline
ME$^\text{a}$  & -0.055  & 0.055 & -0.00009 \\  
MAE$^\text{b}$ & 0.058   & 0.066 & 0.026    \\
SD$^\text{c}$  & 0.048   & 0.055 & 0.034    \\
\hline\hline
\multicolumn{4}{l}{$^\text{a}$Mean error. $^\text{b}$Mean absolute error.}\\
\multicolumn{4}{l}{$^\text{c}$Standard deviation.}\\
\end{tabular*}
\end{table}

\section{Band structure calculations}
\begin{figure}[ht]
 \begin{center}
  \begin{tabular}{cc}
   \includegraphics[width=0.4\textwidth]{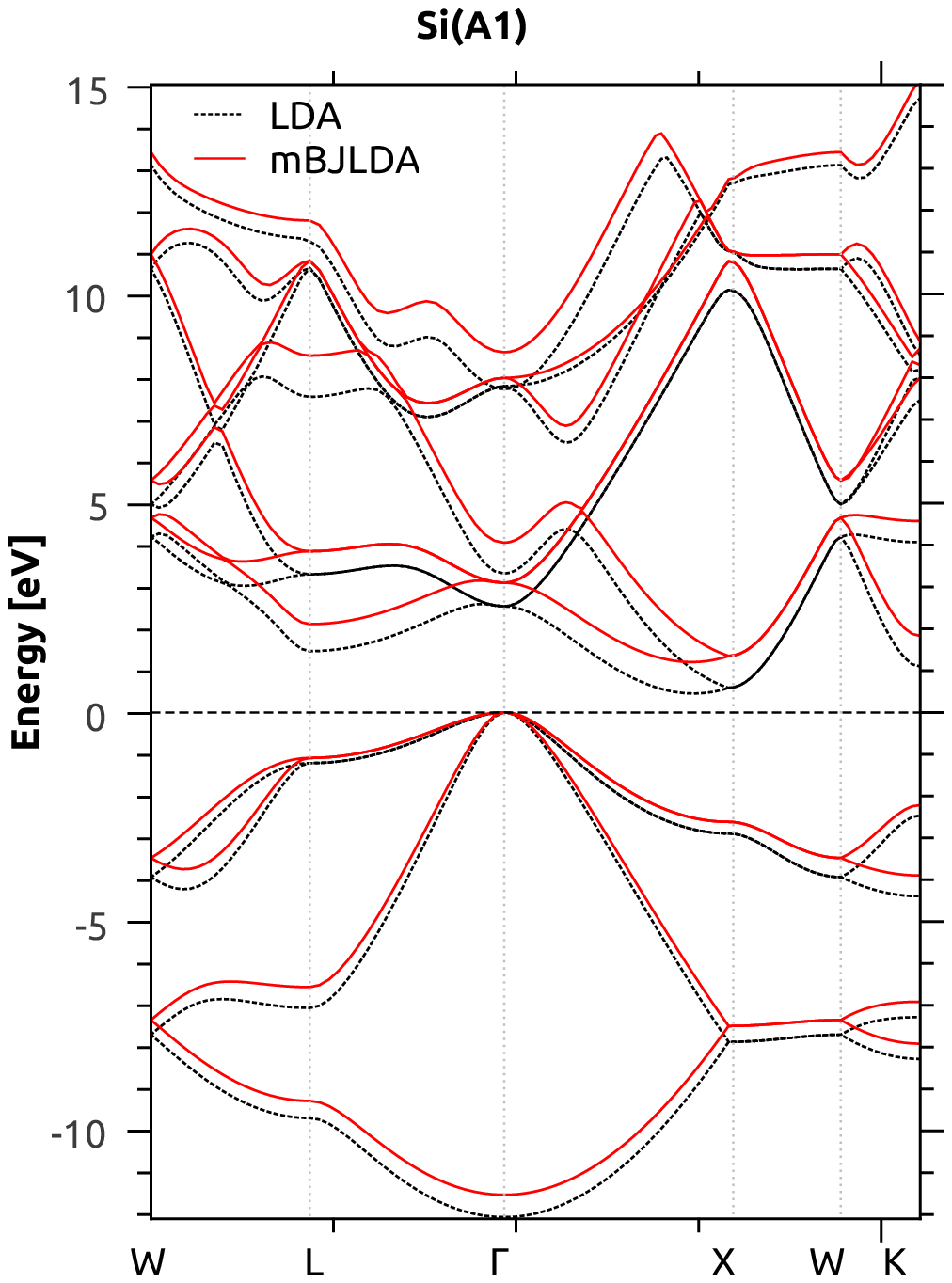} & \includegraphics[width=0.4\textwidth]{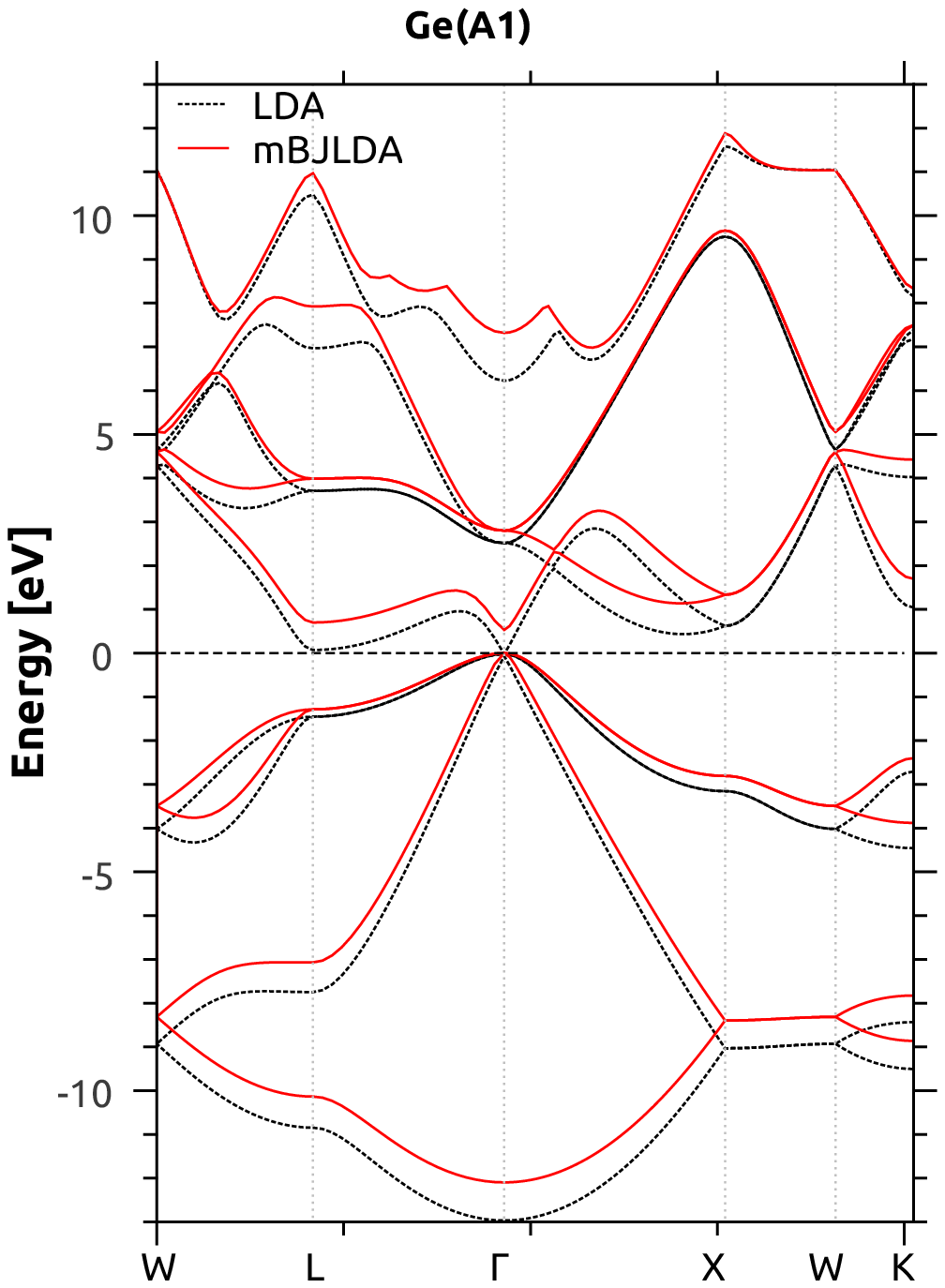} \\
   \includegraphics[width=0.4\textwidth]{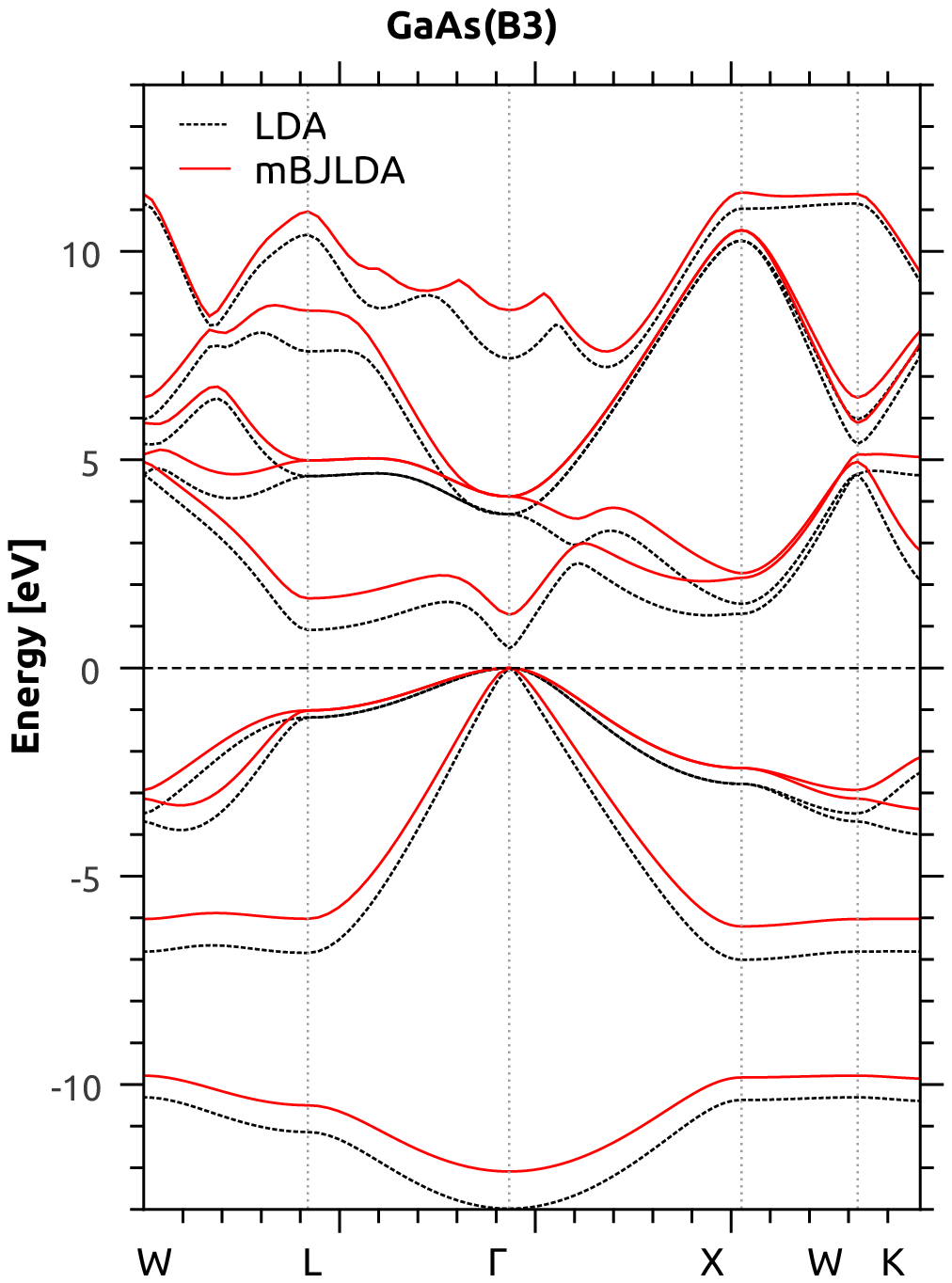} & \includegraphics[width=0.4\textwidth]{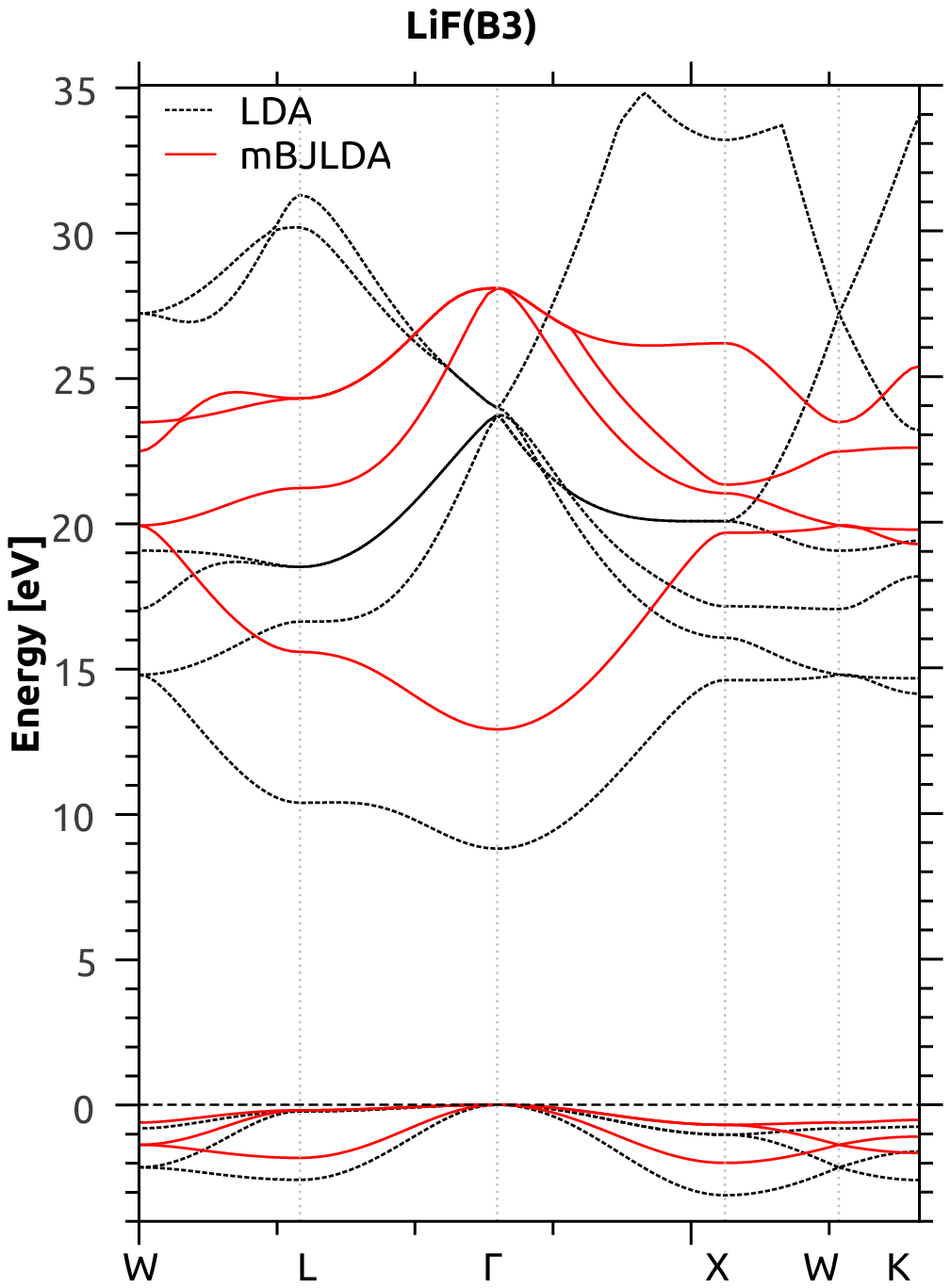} \\
\end{tabular}
 \end{center}
\caption{\label{fig:fig1}The band structure obtained for Si, Ge, GaAs, and LiF with LDA and mBJLDA. The origin is at the top of the valence band. In parenthesis 
the crystal structure is shown.}
\end{figure}

\begin{figure}[ht]
 \begin{center}
   \includegraphics[width=0.4\textwidth]{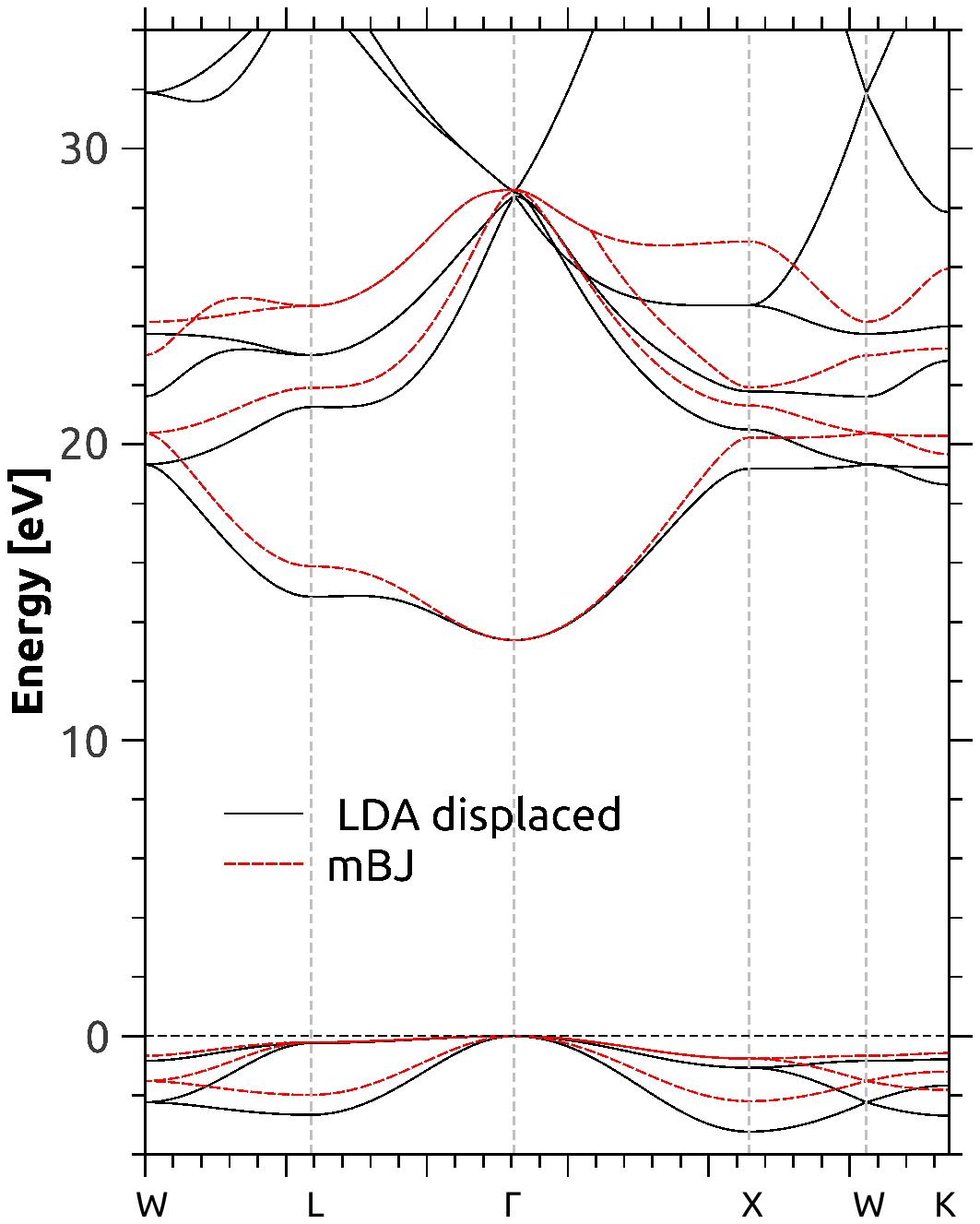}
 \end{center}
\caption{\label{fig:fig2}The band structure obtained for LiF with LDA and mBJLDA.}
\end{figure}

In Fig.\ref{fig:fig1}, we compare the band structure for Si, Ge, GaAs and LiF calculated with the LDA and with the mBJLDA potential. As a general result, this 
potential causes a rigid displacement of the conduction bands toward higher energies with respect to the top of the valence band with small differences in the 
dispersion at some regions of the Brillouin zone but reproducing, in general, the characteristic behavior of the bands for each semiconductor according to 
experiment~\cite{23,24,32,33} as the Wien2k code used to. If we look at the resulting band structures in more detail, we see that a rigid displacement of only 
the conduction bands as to reproduce the mBJLDA predicted gap value will cause some differences in the upper conduction bands although the ones just above the 
upper gap edge will match quite well with the calculation using the mBJLDA potential. This is roughly true for the first three semiconductors presented in 
Fig.\ref{fig:fig1}. LiF deserves a more detailed discussion. In Fig.\ref{fig:fig2}, we show the result of such a displacement. We can see that clear differences 
arise that might influence a calculation of a system containing LiF as one of its elements. So, we conclude that, in general, a rigid 
displacement of a band structure calculation with the old formulation of Wien2K will not be accurate enough for certain purposes as it might occur in the 
calculation of the band structure of certain interfaces containing a semiconductor as one of its elements, for example. The direct use of the previous version
of the Wien2K code in this case, will not be accurate enough as well.\\

In Table \ref{tab:Tabla4}, we show the result of the calculation using the mBJLDA potential with the $a_{LDA}$, $a_{GGA}$ and $a_{Avg}$ to survey the respective 
differences in the gap value as compared to experiment~\cite{23,24,25}. We can see that if we use the lattice parameter from a LDA optimization, mBJ($a_{LDA}$), 
the prediction for the gap deviates by less than 10\% for 21 of the 41 semiconductors considered, between 10-20\% for 9 of them and more then 20\% for the 
remaining 11 (Ge and GaAs among them). Using GGA, we get 19, 14, and 6, respectively. When using the average value of the two, we get 25, 7,and 7 a result that 
improves the calculation at relatively low computational cost. There are, nevertheless, two semiconductors that we did not include in the two last cases just 
presented, MgS(B3) and MgTe. It turns out that both, the calculation using $a_{GGA}$ and $a_{Avg}$ give the wrong result that the gap is indirect while $a_{LDA}$ 
gives rise to a correct direct gap for both semiconductors and predicts its value with less than 10\% deviation in both cases. 
Further, to explore the absolute capacity of the mBJLDA potential to predict the gap value correctly, we have used the experimental lattice parameter value 
at low temperature ($a_{LT}$) where it existed in the literature to perform the calculation. It is interesting that the result is not as good as one would expect 
as it can be seen from Table IV. This calculation shows the best that the mBJLDA potential can do, a result that might be important to bear in mind. It looks 
like that when no data are available, the best result is obtained by taking the average of the lattice parameters resulting from an LDA optimization and a 
GGA one.
Furthermore, it is interesting to observe that small differences in the lattice parameter can give rise to noticeable differences in the predicted value for the gap. 
An example of this is Ge. A small difference in the lattice parameter as 0.5\% (LDA) gives rise to a 23\% deviation of the gap value as compared to experiment. 
Within the GGA optimization, a 1.9\% lattice difference generates a 5.4\% off value for the gap and a 0.6\% difference in the lattice parameter gives an 8.1\% 
difference in the gap when the average is taken. A very interesting issue is that when the experimental value of the lattice parameter is introduced into the 
mBJLDA code, important deviations from experiment occur. In Table \ref{tab:Tabla4}, we have included 14 semiconductors for which we found an experimental low 
temperature lattice parameter value. We get the experimental gap value only for one of them (Si). With less than 10\% deviation we found 7 (C, MgO, AlAs, SiC, 
GaAs, GaP, and CdTe); a deviation between 10-20\% occurs in 3 (Ge, InP, and AlN). Very important deviations from experiment occur in CuCl ( 47.5\%), and CuBr 
(45.3\%). These results are facts to bear in mind for a proper evaluation of the performance of the mBJLDA potential since the quality of the optimization procedure 
is judged from the deviation of the predicted lattice parameter from the experimental value. This judgment implies that the best result for the predicted band gap 
value is obtained when the experimental lattice parameter is introduced into the code. Actually any code should, first of all, be examined in this sense so that 
the selection of an optimization procedure based on the deviation of the predicted lattice parameter from experiment, really makes sense. So, the mBJLDA potential gives 
rise to inconsistencies that emphasise its empirical character. 

\begin{table}[h]
\caption{\label{tab:Tabla4}Calculations of the gap ($E_g$) in eV, from the potential mBJLDA using the parameters in Table \ref{tab:Tabla2}. The lattice 
parameter $a_{Avg}$ is the average value between $a_{LDA}$ and $a_{GGA}$. The experimental values were taken from references~\cite{23,24,25}. The crystal 
structure and percentage difference with respect to experiment is shown in parenthesis.}
     \begin{tabular*}{0.85\textwidth}{@{\extracolsep{\fill}}  l c c c c c }\hline\hline
           & \multicolumn{5}{c}{Gap} \\\hline
Solid   & $E_{g}^{mBJ(a_{LDA})}$ & $E_{g}^{mBJ(a_{GGA})}$ & $E_{g}^{mBJ(a_{LT})}$ & $E_{g}^{mBJ(a_{Avg})}$ &  $E_{g}^{Expt}$  \\\hline
C(A1)      & 4.98 (-9.1\%)   &  4.92 (-10.2\%) & 4.93 (-10.0\%) & 4.95 (-9.7\%)  & 5.48* \\ 
Si(A1)     & 1.13 (-3.4\%)   &  1.20 (2.6\%)   & 1.17 (0.0\%)   & 1.17 (0.0\%)   & 1.17* \\
Ge(A1)     & 0.91 (23\%)     &  0.70 (-5.4\%)  & 0.84 (13.5\%)  & 0.80 (8.1\%)   & 0.74$_{\text {\tiny 1.5 K}}$ \\
MgO(B1)    & 7.57 (-2.6\%)   &  6.88 (-11.5\%) & 7.26 (-6.6\%)  & 7.22 (-7.1\%)  & 7.77$_{\text {\tiny 7.7 K}}$ \\  
AlAs(B3)   & 2.13 (-4.5\%)   &  2.22 (-0.4\%)  & 2.14 (-4.0\%)  & 2.17 (-2.7\%)  & 2.23$_{\text {\tiny 4 K}}$   \\
SiC(B3)    & 2.21 (-8.7\%)   &  2.28 (-5.8\%)  & 2.26 (-6.6\%)  & 2.26 (-6.6\%)  & 2.42$_{\text {\tiny 2 K}}$   \\
AlP(B3)    & 2.28 (-6.9\%)   &  2.37 (-3.3\%)  &        -       & 2.33 (-4.9\%)  & 2.45$_{\text {\tiny 4 K}}$   \\
GaN(B3)    & 3.13 (-10.6\%)  &  2.74 (-21.7\%) &        -       & 2.94 (-16.0\%) & 3.50$_{\text {\tiny 1.6 K}}$ \\
GaAs(B3)   & 1.84 (21.1\%)   &  1.28 (-15.8\%) & 1.62 (6.6\%)   & 1.56 (2.6\%)   & 1.52* \\
InP(B3)    & 1.70 (19.7\%)   &  1.34 (-5.6\%)  & 1.59 (12.0\%)  & 1.52 (7.0\%)   & 1.42$_{\text {\tiny 1.6 K}}$  \\
AlSb(B3)   & 1.76 (4.8\%)    &  1.84 (9.5\%)   &      -         & 1.80 (7.1\%)   & 1.68$_{\text {\tiny 27 K}}$  \\
GaSb(B3)   & 1.20 (46.3\%)   &  0.68 (-17.1\%) &      -         & 0.90 (9.8\%)   & 0.82* \\
GaP(B3)    & 2.18 (-7.2\%)   &  2.30 (-2.1\%)  & 2.24 (-4.7\%)  & 2.24 (-4.7\%)  & 2.35* \\
InAs(B3)   & 0.77 (83.3\%)   &  0.34 (-19.0\%) &       -        & 0.55 (31.0\%)  & 0.42$_{\text {\tiny 4.2 K}}$  \\
InSb(B3)   & 0.59 (145.8\%)  &  0.09 (-62.5\%) &       -        & 0.31 (29.2\%)  & 0.24$_{\text {\tiny 1.8 K}}$  \\
CdS(B3)    & 2.68 (8.1\%)    &  2.52 (1.6\%)   &       -        & 2.61 (5.2\%)   & 2.48$_{\text {\tiny 4.2 K}}$  \\ 
CdTe(B3)   & 1.80 (12.5\%)   &  1.54 (-3.8\%)  & 1.73 (8.1\%)   & 1.67 (4.4\%)   & 1.60* \\
CdSe(B3)   & 1.99 (12.4\%)   &  1.77 (0.0\%)   &       -        & 1.87 (5.6\%)   & 1.77* \\
ZnS(B3)    & 3.85 (1.0\%)    &  3.55 (-6.8\%)  &       -        & 3.70 (-2.9\%)  & 3.81* \\
ZnSe(B3)   & 2.90 (2.8\%)    &  2.56 (-9.2\%)  &       -        & 2.74 (-2.8\%)  & 2.82$_{\text {\tiny 6 K}}$  \\
ZnTe(B3)   & 2.52 (5.4\%)    &  2.19 (-8.4\%)  &       -        & 2.38 (-0.4\%)  & 2.39$_{\text {\tiny 2 K}}$ \\
MgS(B1)    & 4.12 (-8.4\%)   &  4.02 (-10.7\%) &       -        & 4.07 (-9.6\%)  & 4.50$_{\text {\tiny 77 K}}$ \\
MgS(B3)    & 5.18 (-4.1\%)   &  -              &       -        & -              & 5.40$_{\text {\tiny 5 K}}^\text{a}$ \\
MgTe(B3)   & 3.59 (-2.2\%)   &  -              &       -        & -              & 3.67$_{\text {\tiny 2 K}}^\text{b}$ \\
CaO(B1)    & 7.55 (-3.2\%)   & 6.97 (-10.6\%)  &       -        & 7.31 (-6.3\%)  & 7.80*$^+$ \\
CuCl(B3)   & 2.00 (-41\%)    & 1.72 (-49.3\%)  & 1.78 (-47.5\%) & 1.85 (-49.3\%) & 3.39$_{\text {\tiny 2 K}}$  \\
CuBr(B3)   & 1.83 (-40.4\%)  & 1.62 (-47.2\%)  & 1.68 (-45.3\%) & 1.71 (-44.3\%) & 3.07$_{\text {\tiny 1.6 K}}$ \\
AgF(B1)    & 1.80 (-35.7\%)  & 2.44 (-12.9\%)  &       -        & 2.22 (-20.7\%) & 2.80$_{\text {\tiny 4.8 K}}$ \\
AgI(B1)    & 2.72 (-6.5\%)   & 2.83 (-2.7\%)   &       -        & 2.77 (-4.8\%)  & 2.91$_{\text {\tiny 4 K}}$ \\
GaN(B4)    & 3.13 (-5.2\%)   & 3.13 (-5.2\%)   &       -        & 3.13 (-5.2\%)  & 3.30$_{\text {\tiny 10 K}}^\text{c}$ \\
InN(B4)    & 0.82 (15.5\%)   &  0.82 (15.5\%)  &       -        & 0.82 (15.5\%)  & 0.71$_{\text {\tiny 10 K}}^\text{d}$ \\
AlN(B4)    & 5.52 (-10.8\%)  &  5.53 (-10.7\%) & 5.56 (-10.2\%) & 5.53 (-10.7\%) & 6.19$_{\text {\tiny 7 K}}^\text{e}$ \\
ZnO(B4)    & 2.77 (-19.5\%)  &  2.77 (-19.5\%) & 2.72 (-20.9\%) & 2.76 (-19.7\%) & 3.44* \\ 
\hline
LiF(B1)    & 13.80 (1.5\%)   &  12.92 (-5.0\%) &       -        & 13.40 (-1.5\%) & 13.6$_{\text {\tiny 300 K}}^\text{f}$ \\
BP(B3)     & 1.80 (-10.0\%)  &  1.87 (-6.5\%)  &        -       & 1.83 (-8.5\%)  & 2.00$_{\text {\tiny 300 K}}$ \\
BN(B3)     & 5.86 (-5.5\%)   &  5.83 (-6.0\%)  &        -       & 5.85 (-5.6\%)  & 6.20$_{\text {\tiny 300 K}}$ \\
MgSe(B1)   & 2.95 (19.4\%)   &  2.84 (15.0\%)  &       -        & 2.89 (17.0\%)  & 2.47$_{\text {\tiny 300 K}}$ \\
BaS(B1)    & 3.23 (-16.8\%)  & 3.37 (-13.1\%)  &       -        & 3.31 (-14.7\%) & 3.88$_{\text {\tiny 300 K}}$ \\
BaSe(B1)   & 2.80 (\-21.8\%) & 2.94 (-17.9\%)  &       -        & 2.87 (-19.8\%) & 3.58$_{\text {\tiny 300 K}}$ \\
BaTe(B1)   & 2.13 (-30.8\%)  & 2.34 (-24.0\%)  &       -        & 2.24 (-27.3\%) & 3.08$_{\text {\tiny 300 K}}$ \\ 
BAs(B3)    & 1.69 (15.8\%)   &  1.75 (19.9\%)  &       -        & 1.72 (17.8\%)  & 1.46$_{\text {\tiny 300 K}}$ \\
\hline\hline
\multicolumn{6}{l}{*Extrapolated values.$^+$Reported value are direct gap at $\Gamma$.$^\text{a}$Ref~\cite{34}.$^\text{b}$Ref.~\cite{35}.}\\
\multicolumn{6}{l}{$^\text{d}$Ref~\cite{36}.$^\text{e}$Ref~\cite{37}. $^\text{f}$Ref~\cite{38}. $^\text{c}$Ref.~\cite{39}.}
\end{tabular*}
\end{table}  
\newpage
\clearpage

\begin{table}[h]
\caption{\label{tab:Tabla5}Band gap error statistics for compounds in Tables \ref{tab:Tabla4}, in eV. (See text)}
\begin{tabular*}{0.65\textwidth}{@{\extracolsep{\fill}} l c c c c }\hline\hline
\multicolumn{4}{l}{{\small Error relative to LT experiments.}}      \\\hline
               & mBJ($a_{LDA}$) & mBJ($a_{GGA}$) & mBJ($a_{Avg}$) & mBJ($a_{LT}$) \\\hline
ME$^\text{a}$  & -0.15          & -0.32          & -0.24           & -0.87          \\
MAE$^\text{b}$ & 0.33           & 0.35           & 0.30            & 0.45          \\
SD$^\text{c}$  & 0.44           & 0.43           & 0.42            & 1.99          \\\hline
\multicolumn{4}{l}{{\small Error relative to RT experiments.}}\\\hline
               & mBJ(LDA) & mBJ(GGA) & mBJ($a_{Avg}$) &  \\\hline
ME$^\text{a}$  & -0.25   & -0.30 &  -0.27 & \\  
MAE$^\text{b}$ &  0.48   & 0.47  &  0.44  & \\
\hline\hline
\multicolumn{4}{l}{$^\text{a}$Mean error. $^\text{b}$Mean absolute error.} \\
\multicolumn{4}{l}{$^\text{c}$Standard deviation.}
\end{tabular*}
\end{table}

We finally present in Table \ref{tab:Tabla5}, the statistical analysis of the gap value obtained from a band structure calculation with the mBJLDA potential using as
input different lattice parameters obtained from different optimization procedures. The use of the average lattice parameter gives clearly the best result. It is 
worth noting that when the experimental parameter at low temperature, mBJ($a_{LT}$), is introduced in the code to calculate the band gap values, we get, in some cases, 
deviation from the experimental value of the gap that are not expected, in principle.

In a further work~\cite{40} we compare the predicted bsnd gap value as obtained by different methods in the literature. The theoretically well found GW approximation
gives rise to the best predictions. Nevertheless the mBJLDA empirical potential results compare acceptable well.

\section{Conclusions}

We have calculated using the empirical mBJLDA potential~\cite{2}, the band structure of forty one semiconductors and got their band gap value which we 
compared to experiment. In this formulation, there is no expression for the exchange and correlation energy so that the mBJLDA potential can be obtained from 
it, as in the usual theory. Due to this shortcome, no consistent optimization procedure is possible. This is due to the empirical character 
of the mBJLDA potential. As an empirical solution to this problem, Tran and Blaha~\cite{2} suggest the use of a LDA or a GGA optimization to get 
the lattice parameter that goes in a further step into the code that calculates the band structure using the mBJLDA potential. As an overall first sight picture, 
the mBJLDA potential causes a rigid displacement of the conduction band towards higher energies so as to considerably improve the agreement with experiment of the 
band gap values. A closer look reveals that, in some cases, there are noticeable deviations from a rigid displacement of the band structure calculated with the 
previous version of the code that might result in wrong conclusions as it could happen when the band structure of an interface is calculated, for example.
We explored at this point the resulting band gap predicted value as a function of the lattice parameter used. We found two important facts. First, the best result 
for the band gap value is obtained, in general, if the average lattice parameter, $a_{Avg}$, is used ($a_{Avg}=(a_{LDA}+a_{GGA})/2$) where $a_{LDA}$($a_{GGA}$) is 
the lattice parameter that results from a LDA(GGA) optimization. Second, another important result is that if we take the experimental lattice parameter into the code using 
the mBJLDA potential, important deviations from the experimental gap value are obtained, a result to bear in mind for a detailed analysis of the performance of the 
mBJLDA potential. This result is important since it shades the optimization procedure in the sense that a lattice parameter closer to experiment does not 
guarantee a better prediction for the band gap. In spite of the above observations, our work shows that for the forty one semiconductors considered, the mBJLDA
potential represents an important improvement as compared to the results obtained from the previous version of the code. This procedure, all together, only mimics
many-body results. But the mimic is not so bad for reason that do not seem to have theoretical foundation.

In a further work~\cite{40} we have compared the performance of some different methods in the literature to predict the band gap of semiconductors. We find that
the GW approximation (GWA), a theoretical well founded method, gives the best result. The empirical potential in spite of the several problem and shortcomes 
described in this work gives predictions that compare acceptable well. Another factor that might be considered is the computational cost. GWA has a higher 
computational cost. It should be mentioned that this factor looses importance as new computer facilities spread all over the world as it is happening nowadays.
We find, nevertheless, that there are still important issues to be fixed in DFT before we consider the proper prediction of the band gap value of semiconductors, 
a closed problem.


\begin{thebibliography}{}

\bibitem{1} P. Blaha, K. Schwars, G.K.H. Madsen, D. Kvasnicka, and J. Luitz,
{\em WIEN2K:Full Potential-Linearized Augmented Plane waves and Local Orbital Programs 
for Calculating Crystal Properties}, edited by K. Schwars, 
 Vienna University of Technology, Austria, (2001).

\bibitem{2} F. Tran and P. Blaha,
{\em Phys. Rev. Lett.} {\bf 102}, 226401 (2009)

\bibitem{3} G. K. H. Madsen, P. Blaha, K. Schwarz, E. Sjöstedt, and L. ordström, 
{\em Phys. Rev. B} {\bf64}, 195134 (2001).

\bibitem{4} P. Hohenberg and W. Khon,
{\em Phys. Rev.} {\bf 136}, B864 (1964).

\bibitem{5} W. Khon and L.J. Sham,
{\em Phys. Rev.} {\bf 140}, A1133 (1965).

\bibitem{6} J.P. Perdew and Y. Wang,
{\em  Phys. Rev. B} {\bf 45}, 13244 (1992).

\bibitem{7} J.P. Perdew, S. Kurth,  J. Zupan, and P. Blaha,
{\em  Phys. Rev. Lett.} {\bf 82}, 2544 (1999).

\bibitem{8} A.D. Becke and E.R. Johnson,
{\em  J. Chem.} {\bf 124} 221101 (2006).

\bibitem{9} A. D. Becke and M.R. Roussel,
{\em  Phys. Rev. A} {\bf 39}, 3761 (1989).

\bibitem{10}  Murnaghan F.D.,
{\em Am. J. Math.} {\bf 59}, 235 (1937).

\bibitem{11} F. Birch,
{\em Phys. Rev.} {\bf 71} 809 (1947)

\bibitem{12} D. M. Teter, G. V. Gibbs, M. B. Boisen, Jr., D. C. Allan, M. P. Teter,
{\em J. Chem. Phys.} {123}, 174101 (1995).

\bibitem{13} M. L. Cohen,
{\em Phys. Rev. B,} {\bf 32}, 798 (1985).

\bibitem{14} P.E. Van Camp, V.E. Van Doren, and J. T. Devreese,
{\em Phys. Rev. B} {\bf 38}, 12675 (1988).

\bibitem{15} Yingwei Fei,
{\em American Mineralogist} {\bf 84}, 272 (1999).

\bibitem{16} R.A. Miller and Charles S. Smith,
{\em J. of Phys. and Chem. of Sol.} {\bf 25}, 1279 (1964).

\bibitem{17} V. Fiorentini,
{\em Phys. Rev. B} {\bf 46}, 2086 (1992).

\bibitem{18} J. V. Aleksandrov, A. F. Goncharov, S. M. Stishov, and E. Y. Yakovenko, 
{\em JETP Lett.} {\bf 50}, 127 (1989).

\bibitem{19} S-H. Wei and A. Zunger,
{\em Phys. Rev. B} {\bf 60 }, 5404 (1999).

\bibitem{20} D. H. Yean and J. R. Riter,
{\em J. Phys. Chem. Solids}, {\bf 32} 653 (1971).

\bibitem {21} A. Zaoui, and F. El Haj Hassan,
{\em Superlattices and Microstructures}, {\bf 32}, 91 (2002). 

\bibitem{22} H. J. McSkimin, A. Jayaraman. P. Andreatch,
{\em J. Appl. Phys.} {\bf 38}, 2362 (1967).

\bibitem{23} S. Adachi,
{\em  Handbook on Physical Properties of Semiconductors, Vols. I, II and III.} Kluwer Academic Publishers (2004).

\bibitem{24} O. Madelung,
{\em  Data in Sciene and Technology, Semiconductors Group IV Elements and II-V Compounds}. Ed. Springer-Verlag (1991).

\bibitem{25} O. Madelung,
{\em Semiconductors: Data Handbook CD-ROM}. Ed. Springer-Verlag (2003).

\bibitem{26} J. N. Plendl and L. C. Mansur, 
{\em Appl. Opt.} {\bf 11}, 1194 (1972). 

\bibitem{27} A. Taylor and R. M. Jones, {\em in Silicon Carbide, A High Temperature Semiconductor},
edited by J. R. O'Connor and J. Smiltens (Pergamon, London, 1960), p. 1470.

\bibitem{28} J.L. Staudenmann, R. D. Horning, R. D. Knox, D. K. Arch, and J. L. Schmit,
{\em Appl. Phys. Lett.} {\bf 48}, 994 (1986).

\bibitem{29} A. Bouhemadou, R. Khenata, F. Zegrar, M. Sahnoun, H. Baltache and A.H. Reshak,
{\em Comp. Mat. Sc.}, {\bf 2}, 263 (2006).

\bibitem{30} W. M. Vim and R. J. Paff,
{\em J. Appl. Phys.} {\bf45}, 1456 (1974).

\bibitem{31} R. R. Reeber,
{\em J. Appl. Phys.} {\bf 41}, 5063 (1970).

\bibitem{32} N. Wang, M. Rohlfing, P. Kr\"uger, and J. Pollmann,
{\em Phys.Rev. B} {\bf 67}, 115111 (2003).

\bibitem{33} S. Antoci and L. Mihich,
{\em Phys.Rev. B} {\bf 21}, 799 (1980).

\bibitem{34} D. Wolverson, D. M. Bird C. Bradford, K. A. Prior, and B. C. Cavenett,
{\em Phys. Rev. B}, {\bf 64}, 113203 (2001).

\bibitem{35} K. Watanabe, M. Th. Litz, M. Korn, W. Ossau, A. Waag et al.
{\em J. Appl. Phys. } {\bf 81}, 451 (1997).

\bibitem{36} M. Feneberg, J. Daubler, K. Thonke, and R. Sauer.
{\em Phys. Rev. B}, {\bf 77}, 245207 (2008).

\bibitem{37} H.Morkoc,
{\em Handbook of Nitride Semiconductors and Devices}, Ed. Wiley-VCH (2008).

\bibitem{38} D.M. Roessler and W.C. Walker,
{\em J. Phys. Chem. Solids}, {\bf 28}, 1507 (1967).

\bibitem{39} G. Ram\'irez-Flores, H. Navarro-Contreras, A. Lastras-Mart\'inez, R. C. Powell and J. E. Greene,
{\em Phys. Rev. B } {\bf 50}, 8433 (1994). 

\bibitem{40} J. A. Camargo-Mart\'inez and R. Baquero, {\em arXiv:1208.2057v1} [cond-mat.str-el]

\end{thebibliography}
\end{document}